\documentclass[twocolumn,twocolappendix]{aastex631}

\usepackage{bm}
\newcommand{\reviseone}{} % {\reviseone{test.}}
\newcommand{\revisetwo}{} % {\revisetwo{test.}}
\newcommand{\revisethree}{} % {\revisetwo{test.}}
\newcommand{\revisefour}{} % {\revisetwo{test.}}
\newcommand{\revisefive}{} % {\revisetwo{test.}}

\received{December 7, 2022}
\revised{April 27, 2023}
\accepted{June 15, 2023}

\submitjournal{ApJ}

\shorttitle{Compressive Strength of Dust Aggregates} % 39 characters (not more than 44 characters)
\shortauthors{Tatsuuma et al.}
\graphicspath{{./}{figures/}}

\begin{document}

\title{Formulating Compressive Strength of Dust Aggregates from Low to High Volume Filling Factors with Numerical Simulations}
\correspondingauthor{Misako Tatsuuma}
\email{misako.tatsuuma@gmail.com}

\author[0000-0003-1844-5107]{Misako Tatsuuma}
\affiliation{Department of Earth and Planetary Sciences, Tokyo Institute of Technology, 2-12-1 Ookayama, Meguro-ku, Tokyo 152-8551, Japan}
\affiliation{Division of Science, National Astronomical Observatory of Japan, 2-21-1 Osawa, Mitaka, Tokyo 181-8588, Japan}

\author[0000-0003-4562-4119]{Akimasa Kataoka}
\affiliation{Division of Science, National Astronomical Observatory of Japan, 2-21-1 Osawa, Mitaka, Tokyo 181-8588, Japan}

\author[0000-0002-1886-0880]{Satoshi Okuzumi}
\affiliation{Department of Earth and Planetary Sciences, Tokyo Institute of Technology, 2-12-1 Ookayama, Meguro-ku, Tokyo 152-8551, Japan}

\author[0000-0001-9659-658X]{Hidekazu Tanaka}
\affiliation{Astronomical Institute, Tohoku University, 6-3 Aramaki, Aoba-ku, Sendai, Miyagi 980-8578, Japan}

\begin{abstract} % not more than 250 words (250 words)

% background
Compressive strength is a key to understanding the internal structure of dust aggregates in protoplanetary disks and their resultant bodies, such as comets and asteroids in the Solar System.
Previous work has modeled the compressive strength of highly-porous dust aggregates with volume filling factors lower than 0.1.
However, a comprehensive understanding of the compressive strength from low ($<0.1$) to high ($>0.1$) volume filling factors is lacking.
% purpose
In this paper, we investigate the compressive strength of dust aggregates by using aggregate compression simulations resolving constituent grains based on JKR theory to formulate the compressive strength comprehensively.
% method
We perform a series of numerical simulations with moving periodic boundaries mimicking the compression behavior.
% result
As a result, we find that the compressive strength becomes sharply harder when the volume filling factor exceeds 0.1.
% conclusion
We succeed in formulating the compressive strength comprehensively by taking into account the rolling motion of aggregates for low volume filling factors and the closest packing of aggregates for high volume filling factors.
We also find that the dominant compression mechanisms for high volume filling factors are sliding and twisting motions, while rolling motion dominates for low volume filling factors.
We confirm that our results are in good agreement with previous numerical studies.
We suggest that our analytical formula is consistent with the previous experimental results if we assume the surface energy of silicate is $\simeq210\pm90\mathrm{\ mJ\ m^{-2}}$.
Now, we can apply our results to properties of small compact bodies, such as comets, asteroids, and pebbles.

\end{abstract}

% https://astrothesaurus.org

\section{Introduction} \label{sec:intro}

% 1. the first step of planet formation: formation of dust aggregates
The first step of planet formation is the coagulation of (sub)micron-sized dust grains.
The aggregations of dust grains are called dust aggregates {\reviseone{\citep[e.g.,][]{Smirnov1990,Meakin1991,Ossenkopf1993,Dominik1997,Wurm1998,Kempf1999,Blum2000,Krause2004,Paszun2006,Paszun2008,Paszun2009,Wada2007,Wada2008,Wada2009,Wada2013,Suyama2008,Suyama2012,Okuzumi2009dustagg,Geretshauser2010,Geretshauser2011}}}.
{\revisetwo{In the first stage of dust growth, a dust aggregate hits another aggregate and sticks to it.
This process {\revisethree{produces}} fractal aggregates called ballistic cluster-cluster aggregates \citep[BCCAs, e.g.,][]{Mukai1992}}.}
The coagulation of dust aggregates leads to the formation of planetesimals, which are kilometer-sized building blocks of planets {\reviseone{\citep[e.g.,][]{Okuzumi2012,Kataoka2013L}}}.
{\revisetwo{There is another scenario that dust aggregates grow {\revisethree{into}} millimeter-sized compact pebbles, and pebbles coagulate to form planetesimals by some instabilities or collisions \citep[e.g.,][]{Johansen2007,Windmark2012b,Davidsson2016,WahlbergJansson2017,Yang2017,Lorek2018,Fulle2019}.}}
{\revisethree{In this scenario, planetesimals are pebble aggregates whose internal structure is different from dust aggregates in this work.}}

% 2. important in growth process of dust aggregates: compression (because difficult)
{\reviseone{Compression of dust aggregates is a key process during their growth.
There are several compression mechanisms: collisional, disk gas, and self-gravity compression.
Some numerical studies have shown that collisional compression is not sufficient and aggregates' internal densities remain $\sim10^{-5}\mathrm{\ g\ cm^{-3}}$ \citep[e.g.,][]{Okuzumi2012,Kataoka2013L}.
Some experimental studies suggest that the bouncing of dust aggregates leads to compaction \citep[e.g.,][]{Krijt2018}, but numerical simulation studies suggest that the bouncing of highly porous dust aggregates hardly occurs \citep[e.g.,][]{Wada2011}.
As for disk gas and self-gravity compression, the compressive strength of dust aggregates determines their internal densities \citep[e.g.,][]{Blum2004,Paszun2008,Guttler2009,Seizinger2012,Kataoka2013,Kataoka2013L,Omura2017}
The compressive strength also determines the internal structures of larger bodies, such as planetesimals, asteroids, and comets \citep[e.g.,][]{Omura2018,Omura2021}.}}

% 3. what is known about compressive strength: low density formulation
{\reviseone{\citet{Kataoka2013} have modeled the compressive strength of highly-porous dust aggregates with volume filling factors lower than 0.1.
They have analytically formulated the compressive strength by using the volume filling factor and several material parameters, such as monomer (constituent grain) radius and surface energy.}}

% 4. problem in compressive strength: no comprehensive understanding from low to high density
However, a comprehensive understanding of the compressive strength from low ($<0.1$) to high ($>0.1$) volume filling factors is lacking.
{\revisetwo{The compressive strength for high volume filling factors is necessary for applications to comets, asteroids, and pebbles, while for low volume filling factors is necessary for dust growth.}}
Some studies investigated the compressive strength for volume filling factors above 0.1 \citep[e.g.,][]{Blum2004,Paszun2008,Guttler2009,Seizinger2012,Omura2017,Omura2018}.
However, the dependences on material parameters are still unclear and there is a discrepancy between low and high volume filling factors.

% 5. what is done in this paper
In this work, we {\revisetwo{perform numerical simulations of the compression of dust aggregates and formulate the compressive strength that can treat a full range of volume filling factors.}}
We use the same simulation code as \citet{Kataoka2013}, {\reviseone{but we calculate the compressive strength to high volume filling factors to apply it to small bodies in the Solar System with volume filling factors higher than 0.1.
We also investigate the dependences on material parameters, such as monomer radius and surface energy.}}
Finally, we construct a {\revisethree{corrected}} {\revisetwo{analytical formula}} of the compressive strength of dust aggregates based on a simple model to apply it to other parameters.

% 6. organization of this paper
This paper is organized as follows.
In Section \ref{sec:setting}, we explain our simulation settings and {\reviseone{a monomer interaction}} model based on \citet{Dominik1997} and \citet{Wada2007}.
Our simulation settings, such as initial conditions, boundary conditions, and calculation of compressive strength, are the same as those of \citet{Kataoka2013}.
In Section \ref{sec:result}, we show our results of numerical simulations to derive the compressive strength of dust aggregates.
We show fiducial runs, and then we investigate the parameter {\reviseone{dependences}}.
In Section \ref{sec:discuss}, we discuss {\reviseone{parameter dependences and}} the physics behind the compression of dust aggregates.
We show a {\revisethree{corrected}} analytical formula of compressive strength {\reviseone{and}} energy dissipation mechanisms during the compression.
{\reviseone{Then, we}} compare our results with previous experimental and numerical studies {\reviseone{to confirm the validity of our results and discuss interpretations of previous results}}.
Finally, we conclude our work in Section \ref{sec:conclusion}.

\section{Simulation Settings} \label{sec:setting}

In this section, we explain our simulation settings.
First, we introduce {\reviseone{a}} monomer interaction model based on \citet{Dominik1997} and \citet{Wada2007} in Section \ref{subsec:setting:model}.
We also explain {\reviseone{an artificial normal}} damping force.
Second, we describe the {\revisetwo{outline of our simulations}}, where we use periodic boundaries and move them to {\revisetwo{calculate}} the compressive strength in Section \ref{subsec:setting:boundary}.
We also explain the initial conditions {\revisetwo{and the velocity at the {\revisefive{computational}} boundaries}}.
Third, we explain the method to {\revisetwo{calculate}} the compressive strength {\reviseone{and volume filling factor}} in Section \ref{subsec:setting:measure}.

\subsection{Monomer Interaction Model} \label{subsec:setting:model}

% overview of monomer interaction model
We calculate the interactions of spherical monomers in contact by using {\reviseone{a}} theoretical model of \citet{Dominik1997} and \citet{Wada2007} {\revisetwo{based on JKR theory \citep{Johnson1971}}}.
There are four kinds of interactions in this model: normal direction, sliding, rolling, and twisting motions.
The material parameters that are needed to describe the model are the monomer radius $r_0$, material density $\rho_0$, surface energy $\gamma$, Poisson's ratio $\nu$, Young's modulus $E$, and the critical rolling displacement $\xi_\mathrm{crit}$.
{\reviseone{We list the material parameters}} of ice and silicate in Table \ref{tab:parameters}.
{\reviseone{We set the same values to compare our results with those of \citet{Kataoka2013}.}}

% Eroll
{\revisefive{We explain the rolling behavior of two monomers in contact as a consequence of rolling motion dominating during the compression of dust aggregates with volume filling factors lower than 0.1 \citep{Kataoka2013}.}}
Two monomers {\reviseone{roll}} irreversibly after the absolute value of the rolling displacement exceeds the critical limit $\xi_\mathrm{crit}$.
The critical rolling displacement has different values between the theoretical one \citep[$\xi_\mathrm{crit}=2$ \AA,][]{Dominik1997} and the experimental one \citep[$\xi_\mathrm{crit}=32$ \AA,][]{Heim1999}.
We adopt $\xi_\mathrm{crit}=8$ {\AA} as a fiducial value of ice according to \citet{Kataoka2013} and investigate the dependence of our results on $\xi_\mathrm{crit}$ in Section \ref{subsec:result:parameter}.
The energy needed for a monomer to roll a distance of $(\pi/2)r_0$ is given as
\begin{eqnarray}
E_\mathrm{roll} &=& 6\pi^2\gamma r_0\xi_\mathrm{crit}\nonumber\\
&\simeq& 4.7\times10^{-16}\mathrm{\ J}\nonumber\\
&&\times\left(\frac{\gamma}{100\mathrm{\ mJ\ m^{-2}}}\right)\left(\frac{r_0}{0.1\mathrm{\ \mu m}}\right)\left(\frac{\xi_\mathrm{crit}}{8\textrm{\ \AA}}\right).
\label{eq:E_roll}
\end{eqnarray}
{\revisetwo{For details, see Sections 2.2.2 and 3 of \citet{Wada2007}.}}

% damping
{\revisefive{We add an artificial normal damping force proportional to a dimensionless damping coefficient $k_\mathrm{n}$.
For details, see Section 2.2 of \citet{Tatsuuma2019}.}}
The force in the normal direction induces oscillations of two monomers in contact.
In reality, the oscillations would {\reviseone{decay}} because of viscoelasticity or hysteresis of monomers \citep[e.g.,][]{Greenwood2006,Tanaka2012,Krijt2013}.
We adopt $k_\mathrm{n}=0.01$ according to \citet{Kataoka2013}, although they have revealed that the damping coefficient does not change the compressive strength.

\begin{deluxetable*}{lccccc}
\tablecaption{Material parameters of ice and silicate
\label{tab:parameters}}
\tablewidth{0pt}
\tablehead{
\colhead{Parameter} & \colhead{Ice 0.1 $\mathrm{\mu m}$ (fiducial)} & \colhead{Ice (others)} & \colhead{Ice 1.0 $\mathrm{\mu m}$} & \colhead{Silicate 0.1 $\mathrm{\mu m}$} & \colhead{Silicate 1.0 $\mathrm{\mu m}$}
}
\startdata
Monomer radius $r_0$ ($\mathrm{\mu m}$) & 0.1 & 0.1 & 1.0 & 0.1 & 1.0 \\
Material density $\rho_0$ (g cm$^{-3}$) & 1.0 & 1.0 & 1.0 & 2.65 & 2.65 \\
Surface energy $\gamma$ (mJ m$^{-2}$) & 100 & 100 & 100 & 20 & 20 \\
Poisson's ratio $\nu$ & 0.25 & 0.25 & 0.25 & 0.17 & 0.17 \\
Young's modulus $E$ (GPa) & 7 & 7 & 7 & 54 & 54 \\
Critical rolling displacement $\xi_\mathrm{crit}$ (\AA) & 8 & 2, 4, 16, 32 & 8 & 20 & 20 \\
The number of monomers $N$ & 16384 & 16384 & 16384 & 16384 & 16384 \\
Strain rate parameter $C_\mathrm{v}$ & $1\times10^{-7}$ & $3\times10^{-7}$ & $1\times10^{-7}$ & $1\times10^{-7}$ & $1\times10^{-7}$ \\
Damping coefficient $k_\mathrm{n}$ & 0.01 & 0.01 & 0.01 & 0.01 & 0.01 \\
\enddata
\tablecomments{We use $\rho_0$, $\gamma$, $\nu$, $E$, and $\xi_\mathrm{crit}$ of ice {\revisetwo{of}} \citet{Israelachvili1992} and \citet{Dominik1997} and of silicate {\revisetwo{of}} \citet{Seizinger2012}.}
\end{deluxetable*}

\subsection{Compression Simulation Setups} \label{subsec:setting:boundary}

\begin{figure*}[t!]
\plotone{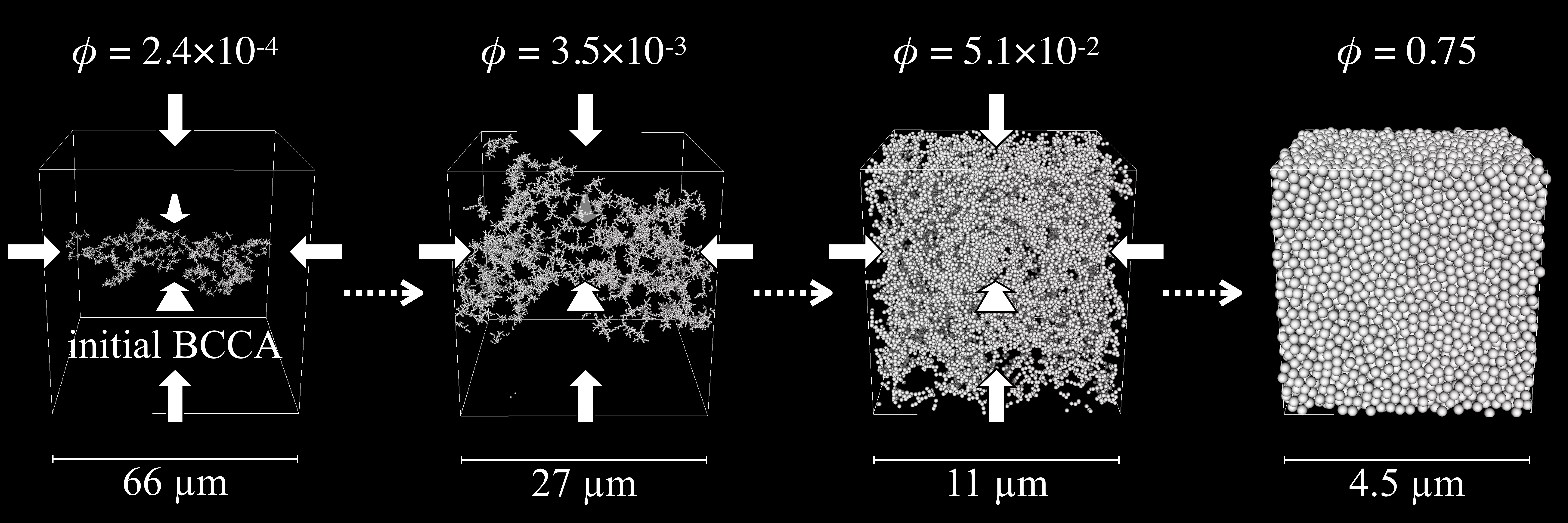}
\caption{{\reviseone{Outline}} of our simulations.
Each aggregate contains 16384 ice monomers {\reviseone{with a}} radius {\reviseone{of}} 0.1 $\mathrm{\mu m}$.
Each white cube shows the {\revisetwo{computational}} box with periodic boundaries.
We compress it three-dimensionally as shown by the white allows.
We put each volume filling factor and size of the {\revisetwo{computational}} box above and below each snapshot, respectively.
{\revisetwo{In our simulations,}} the final volume filling factor {\revisetwo{can exceed}} a value {\revisetwo{for}} the closest packing $\sqrt{2}\pi/6=0.74$ because monomers can deform elastically.
\label{fig:snapshot}}
\end{figure*}

% simulation overview
The {\reviseone{outline}} of our numerical simulations is as follows {\reviseone{(see also Figure \ref{fig:snapshot})}}.
First, we randomly create a BCCA.
Second, we compress it {\reviseone{sufficiently slowly}} and isotropically by moving periodic boundaries.
For details of the periodic boundary conditions, see Section 2.3 {\revisetwo{of}} \citet{Kataoka2013}.

% boundary velocity
The velocity at the {\revisefive{computational}} boundaries is {\revisefour{given as
\begin{equation}
v_\mathrm{b} = -\frac{C_\mathrm{v}}{t_\mathrm{c}}L,
\end{equation}
where $C_\mathrm{v}$ is a constant dimensionless strain-rate parameter, $t_\mathrm{c}$ is the characteristic time \citep{Wada2007}, and $L$ is the length of the computational box.
The characteristic time is given as
\begin{eqnarray}
t_\mathrm{c} &=& 0.95\frac{r_0^{7/6}\rho_0^{1/2}}{\gamma^{1/6}E^{\ast1/3}}\nonumber\\
&\simeq&1.9\times10^{-10}\mathrm{\ s}\left(\frac{\gamma}{100\mathrm{\ mJ\ m^{-2}}}\right)^{-1/6}\left(\frac{r_0}{0.1\mathrm{\ \mu m}}\right)^{7/6}\nonumber\\
&&\times\left(\frac{\rho_0}{1\mathrm{\ g\ cm^{-3}}}\right)^{1/2}\left(\frac{E^\ast}{3.73\mathrm{\ GPa}}\right)^{-1/3},
\end{eqnarray}
where $E^\ast$ is the reduced Young's modulus of Young's moduli $E_1$ and $E_2$ defined as
\begin{equation}
\frac{1}{E^\ast}=\frac{1-\nu_1^2}{E_1}+\frac{1-\nu_2^2}{E_2}.
\end{equation}
Here, we assume $E_1=E_2=E$ and $\nu_1=\nu_2=\nu$, and therefore $E^\ast=E/[2(1-\nu^2)]$.
We can also describe the absolute value of the velocity at the {\revisefive{computational}} boundaries as
\begin{eqnarray}
|v_\mathrm{b}| &\simeq& 0.21\mathrm{\ cm\ s^{-1}}\left(\frac{\gamma}{100\mathrm{\ mJ\ m^{-2}}}\right)^{1/6}\left(\frac{r_0}{0.1\mathrm{\ \mu m}}\right)^{-1/6}\nonumber\\
&&\times\left(\frac{\rho_0}{1\mathrm{\ g\ cm^{-3}}}\right)^{-1/2}\left(\frac{E^\ast}{3.73\mathrm{\ GPa}}\right)^{1/3}\left(\frac{C_\mathrm{v}}{1\times10^{-7}}\right)\nonumber\\
&&\times\left(\frac{N}{16384}\right)^{1/3}\phi^{-1/3},
\label{eq:boundary}
\end{eqnarray}
where $N$ is the number of monomers and $\phi$ is the volume filling factor.}}

We adopt $C_\mathrm{v}=1\times10^{-7}$ as a fiducial value because the larger $C_\mathrm{v}$, the higher pressure {\revisetwo{we need}} for the compression of low-density dust aggregates \citep{Kataoka2013}.
For other parameter sets, we adopt $C_\mathrm{v}=3\times10^{-7}$ because simulations of low $C_\mathrm{v}$ are time-consuming.

\subsection{Compressive Strength and Volume Filling Factor Measurements} \label{subsec:setting:measure}

% calculation of compressive strength
We calculate the compressive strength {\revisetwo{$P_\mathrm{calc}$}} in the way described in Section 2.4 {\revisetwo{of}} \citet{Kataoka2013}.
We calculate the translational kinetic energy per unit volume and the sum of the forces acting at all connections per unit volume {\revisetwo{as
\begin{equation}
P_\mathrm{calc} = \frac{2K}{3V} + \frac{1}{3V}\left\langle\sum_{i<j}(\bm{x}_i-\bm{x}_j)\cdot\bm{f}_{i,j}\right\rangle_t,
\label{eq:Pcalc}
\end{equation}
where $V$ is the volume of the computational box, $K$ is the time-averaged kinematic energy of all monomers given as
\begin{equation}
K = \left\langle\sum_{i=1}^N\frac{m_0}{2}\left(\frac{\mathrm{d}\bm{x}_i}{\mathrm{d}t}\right)^2\right\rangle_t,
\end{equation}
$m_0$ is the monomer mass, $\bm{x}_i$ is the coordinates of monomer $i$, $\langle\rangle_t$ is a long-time average, and $\bm{f}_{i,j}$ is the force from monomer $j$ on monomer $i$.
For details of the derivation of compressive strength, see Appendix \ref{apsec:compstrength}.
In our simulations, the second term on the right-hand side of Equation (\ref{eq:Pcalc}) dominates the compressive strength.}}

{\revisetwo{We also calculate the volume filling factor of dust aggregates as
\begin{equation}
\phi_\mathrm{calc} = \frac{(4/3)\pi r_0^3N}{V}.
\label{eq:phi_calc}
\end{equation}}}

We take an average of the compressive strength $P_\mathrm{calc}$ (Equation (\ref{eq:Pcalc})) and the volume filling factor $\phi_\mathrm{calc}$ {\revisetwo{(Equation (\ref{eq:phi_calc}))}} for every 10,000 time-steps.
One time-step in our simulations {\revisefour{is $0.1t_\mathrm{c}=1.9\times10^{-11}$ s}} and 10,000 time-steps correspond to $1.9\times10^{-7}$ s.

\section{Numerical Results} \label{sec:result}

In this section, we report the results of numerical simulations to derive the compressive strength of dust aggregates.
We perform 10 simulations with different BCCAs and take an average of them for every parameter set {\revisefour{to reduce the effect of different monomer configurations}}.
First, we show the results of fiducial runs in Section \ref{subsec:result:fiducial}.
Then, we investigate the parameter {\reviseone{dependences}} in Section \ref{subsec:result:parameter}.
\citet{Kataoka2013} confirmed that the results do not depend on any numerical parameters: the number of monomers $N$, the strain rate parameter $C_\mathrm{v}$, and the damping coefficient $k_\mathrm{n}$.
{\revisefour{For the dependence on the numerical parameters, see Appendix \ref{apsec:parameterdepend}.}}

\subsection{Fiducial Runs} \label{subsec:result:fiducial}

\begin{figure}[ht!]
\plotone{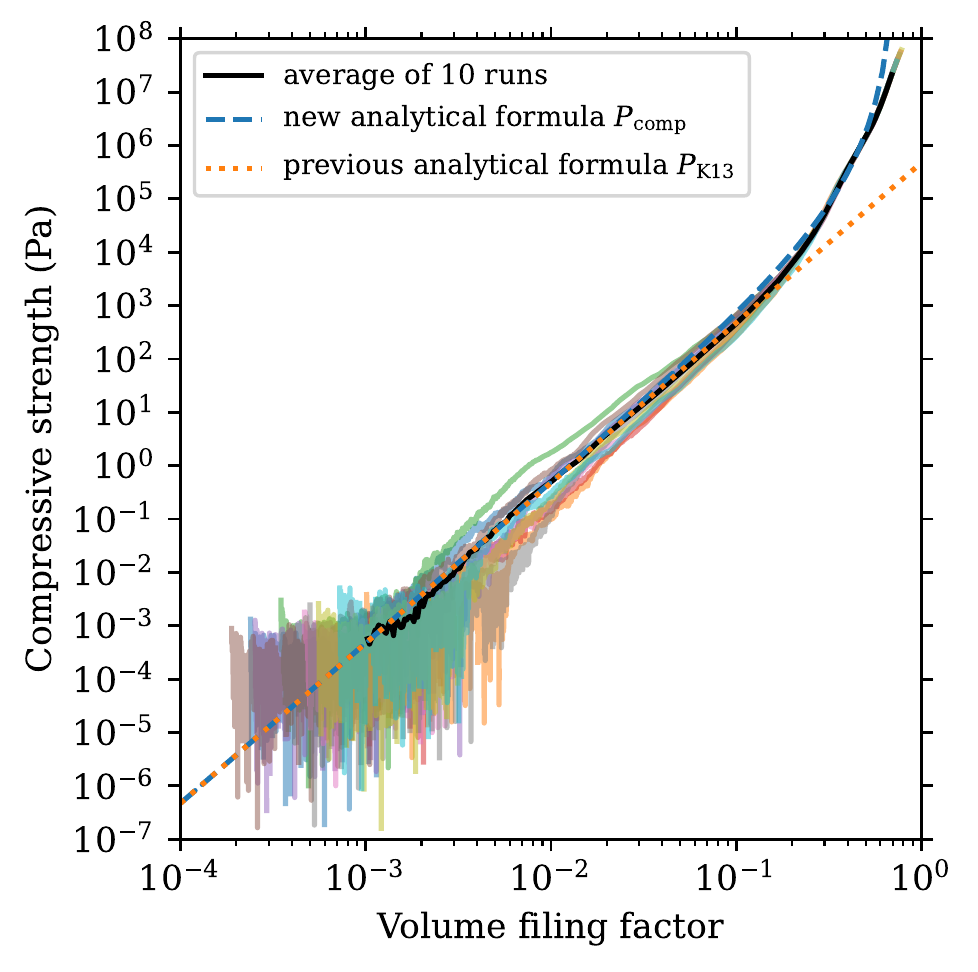}
\caption{Compressive strength against volume filling factor of dust aggregates that contain ice monomers of 0.1-$\mathrm{\mu m}$ radius.
The colored {\revisefour{solid}} lines show the results for {\revisetwo{10}} different initial BCCAs.
The black solid line shows the average of the 10 runs.
{\revisefour{The dashed line shows our corrected analytical formula (Equation (\ref{eq:Pcomp_mod})).
The dotted line shows the analytical formula (Equation (\ref{eq:Pcomp_kataoka})) of \citet{Kataoka2013}.}}
\label{fig:comp_fiducial}}
\end{figure}

% fiducial results
{\revisetwo{Figure \ref{fig:comp_fiducial} shows}} the compressive strength of 10 runs and an average of them for the fiducial parameter set.
The {\revisefour{dotted}} line shows the analytical formula {\revisetwo{of}} \citet{Kataoka2013}{\revisetwo{,}}
\begin{eqnarray}
P_\mathrm{K13} &=& \frac{E_\mathrm{roll}}{r_0^3}\phi^3 \nonumber\\
&\simeq& 4.7\times10^5\mathrm{\ Pa} \nonumber\\
&&\times\left(\frac{\gamma}{100\mathrm{\ mJ\ m^{-2}}}\right)\left(\frac{r_0}{0.1\mathrm{\ \mu m}}\right)^{-2}\left(\frac{\xi_\mathrm{crit}}{8\textrm{\ \AA}}\right)\phi^3.
\label{eq:Pcomp_kataoka}
\end{eqnarray}
Our averaged simulation result is consistent with Equation (\ref{eq:Pcomp_kataoka}) when $\phi\lesssim0.1${\revisefive{.
However, we find that the measured compressive strength for $\phi>0.1$ is significantly higher than predicted from Equation (\ref{eq:Pcomp_kataoka}).}}
{\revisefour{We note that the compressive strength measured in each run has a large scatter.}}

\subsection{Parameter Dependences} \label{subsec:result:parameter}

\begin{figure}[ht!]
\plotone{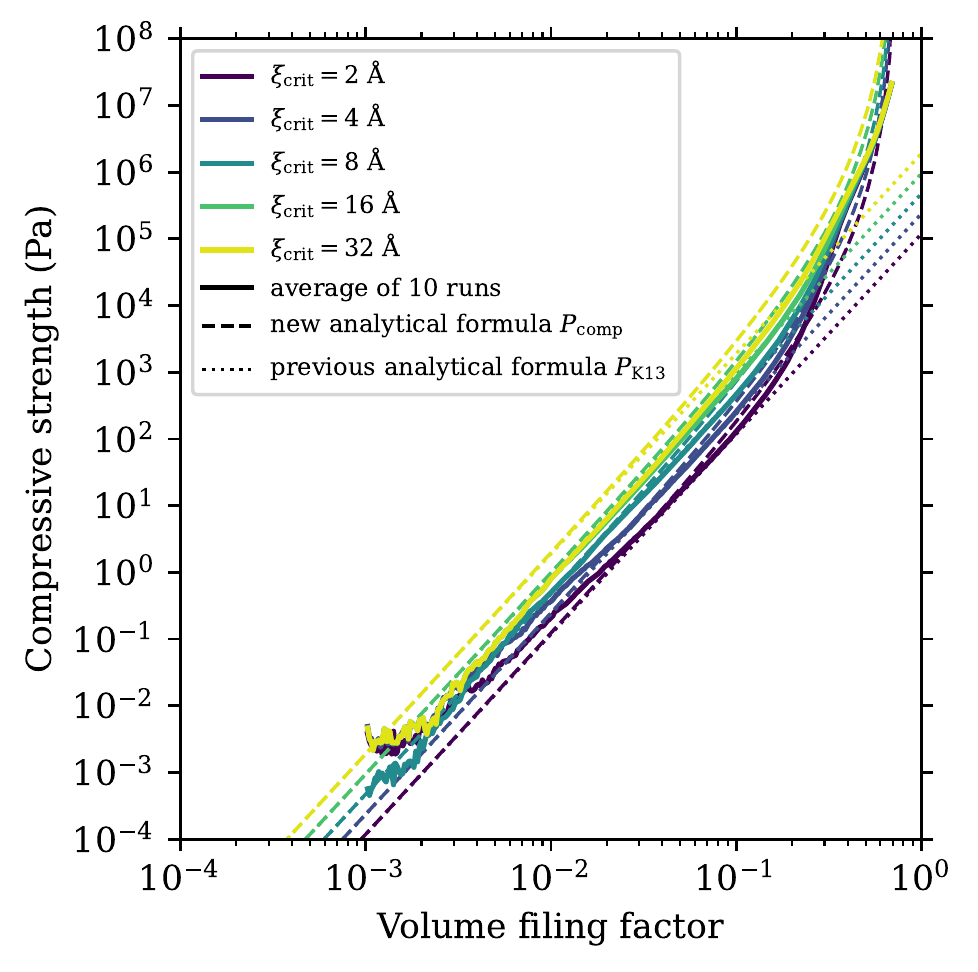}
\caption{Compressive strength against volume filling factor of dust aggregates that contain ice monomers of 0.1-$\mathrm{\mu m}$ radius with different critical rolling displacements $\xi_\mathrm{crit}$.
{\reviseone{The critical rolling displacements are $\xi_\mathrm{crit}=2$, 4, 8, 16, and 32 {\AA} from dark to light colors.}}
We adopt $C_\mathrm{v}=3\times10^{-7}$ for $\xi_\mathrm{crit}=2$, 4, 16, and 32 {\AA} because of the calculation cost.
The other parameters are fiducial values in Table \ref{tab:parameters}.
{\revisefour{The solid, dashed, and dotted lines show the averages of the 10 runs, our corrected analytical formula (Equation (\ref{eq:Pcomp_mod})), and the analytical formula (Equation (\ref{eq:Pcomp_kataoka})) of \citet{Kataoka2013}, respectively.}}
\label{fig:comp_roll}}
\end{figure}

% critical rolling displacement
{\revisefive{We find that the compressive strength does not depend on the critical rolling displacement $\xi_\mathrm{crit}$ when $\phi\gtrsim0.3$.
This is shown in Figure \ref{fig:comp_roll}, where we plot compressive strength when $\xi_\mathrm{crit}=2$, 4, 8, 16, and $32\textrm{\ \AA}$.}}
{\revisefive{In contrast,}} when $\phi\lesssim0.3$, the compressive strength has a dependence on $\xi_\mathrm{crit}$ because the {\reviseone{dominant}} mechanism of {\reviseone{energy dissipation}} is rolling motion.
An exception is when $\xi_\mathrm{crit}=32\textrm{\ \AA}$, {\reviseone{for which the compressive strength curve is nearly identical to that for}} $\xi_\mathrm{crit}=16\textrm{\ \AA}$.
{\reviseone{This is}} because the {\reviseone{dominant}} mechanism of {\reviseone{energy dissipation for $\xi_\mathrm{crit}>16\textrm{\ \AA}$}} is twisting motion \citep{Kataoka2013}.
{\revisefour{We note that the difference in compressive strength due to $\xi_\mathrm{crit}$ is about the same as the difference per run.}}

\begin{figure}[ht!]
\plotone{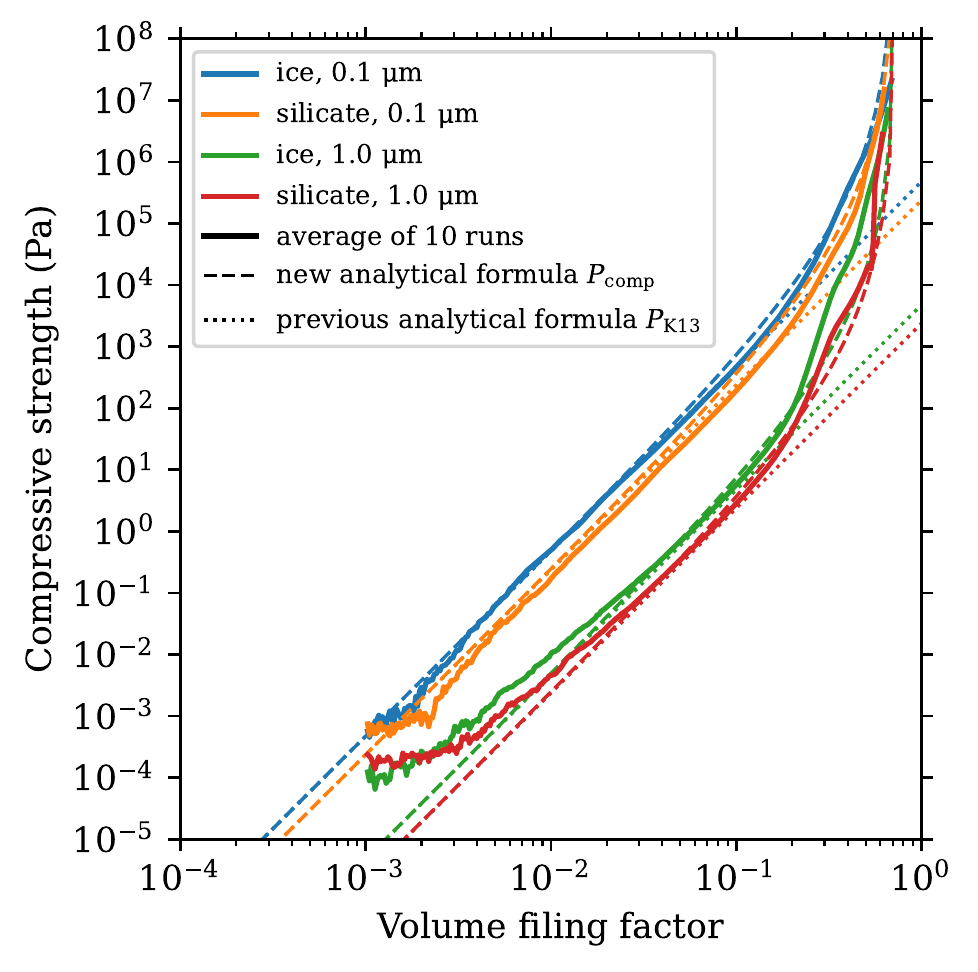}
\caption{Compressive strength against volume filling factor of dust aggregates with different monomer radii and materials.
The {\revisefour{colors}} represent 0.1-$\mathrm{\mu m}$-radius ice (blue), 1.0-$\mathrm{\mu m}$-radius ice (green), 0.1-$\mathrm{\mu m}$-radius silicate (orange), and 1.0-$\mathrm{\mu m}$-radius silicate (red).
The parameters are listed in Table \ref{tab:parameters}.
{\revisefour{The solid, dashed, and dotted lines show the averages of the 10 runs, our corrected analytical formula (Equation (\ref{eq:Pcomp_mod})), and the analytical formula (Equation (\ref{eq:Pcomp_kataoka})) of \citet{Kataoka2013}, respectively.}}
\label{fig:comp_material}}
\end{figure}

% monomer and material parameters
{\revisefive{As for the dependences on the other material parameters, we find that the scaling predicted from Equation (\ref{eq:Pcomp_kataoka}) that the compressive strength scales as $P\propto\gamma\xi_\mathrm{crit}r_0^{-2}$ no longer applies for $\phi>0.1$.
This is shown in Figure \ref{fig:comp_material}, where we plot the compressive strength in the four cases (ice $0.1\mathrm{\ \mu m}$, ice $1.0\mathrm{\ \mu m}$, silicate $0.1\mathrm{\ \mu m}$, and silicate $1.0\mathrm{\ \mu m}$) listed in Table \ref{tab:parameters}.}}
The {\revisefour{dotted}} lines show the analytical formula (Equation (\ref{eq:Pcomp_kataoka})) {\revisetwo{of}} \citet{Kataoka2013}.
{\revisefive{In contrast, for $\phi\lesssim0.1$, our results are consistent with the prediction from Equation (\ref{eq:Pcomp_kataoka}).}}
We note that there are fluctuations of lines when $\phi<10^{-2}$ because dust aggregates are not attached to all {\revisefive{computational}} boundaries.

\section{Discussions} \label{sec:discuss}

In this section, we discuss {\reviseone{parameter dependences and}} the physics behind the compressive strength of dust aggregates.
First, we {\revisethree{correct}} the analytical formula of compressive strength {\reviseone{with}} volume filling {\reviseone{factors}} higher than 0.1 in Section \ref{subsec:discuss:formulate}.
{\revisefour{Second, we discuss the physical validity of the formulated compressive strength in terms of monomer disruption in Section \ref{subsec:discuss:monomerdisruption}.}}
{\revisefour{Third}}, we show the energy dissipation mechanisms during the compression in Section \ref{subsec:discuss:energy}.
Finally, we compare our results with previous experimental and numerical studies in Section \ref{subsec:discuss:compare} {\reviseone{to confirm the validity of our results and discuss interpretations of experimental results}}.

\subsection{Corrected Formula of Compressive Strength} \label{subsec:discuss:formulate}

We have shown in Section \ref{sec:result} that the simple formula of \citet{Kataoka2013} (Equation (\ref{eq:Pcomp_kataoka})) underestimates the compressive strength at $\phi>0.1$.
Here, we propose a corrected formula applicable to both low and high volume filling factors.

The reason why the previous formula is inaccurate for high volume filling factors is that it neglects the finite {\revisefive{volume}} of monomers.
In this regard, the previous formula is similar to the equation of state for ideal gases, {\revisefive{in which the pressure neglects the volume occupied by molecules}}.
As is well known, the ideal gas law breaks down at high densities where the {\revisefive{inter-molecular volume is small compared to the volume occupied by the molecules}}.
Van der Waals' equation of state for real gases takes the finite {\revisefive{volume}} of the molecules into account by simply subtracting the excluded volume from the volume in the ideal gas law.
We expect that a similar correction should improve the accuracy of the previous compressive strength formula.

The {\revisethree{correction}} is as follows.
First, we {\revisefour{invert}} Equation (\ref{eq:Pcomp_kataoka}) as
\begin{equation}
P_\mathrm{\revisefour{comp}} = \frac{E_\mathrm{roll}}{r_0^3}{\revisefour{\phi'}}^3 = \frac{E_\mathrm{roll}}{r_0^3}\left(\frac{NV_0}{V'}\right)^3,
\label{eq:Pcomp_kataoka_mod}
\end{equation}
where $V_0=(4/3)\pi r_0^3$ is the volume of a monomer.
Here, we assume that $V'$ is the volume of the void of dust aggregates, not the volume of dust aggregates.
The volume of the void is almost the same as the volume of dust aggregates when $\phi\lesssim0.1$, while there is a difference between them when $\phi>0.1$.
Second, we determine the excluded volume that cannot be used for compression.
The volume of all monomers $NV_0$ is the excluded volume.
In addition, the void of the closest packed aggregates {\revisefour{$V_\mathrm{cp}-NV_0$}} is the excluded volume{\revisefour{, where $V_\mathrm{cp}$ is the volume of the closest packed aggregates}}.
Therefore, we determine the excluded volume as {\revisefour{$NV_0+V_\mathrm{cp}-NV_0=V_\mathrm{cp}$.}}
{\revisefour{By using the volume filling factor of the closest packed aggregates $\phi_\mathrm{max}=NV_0/V_\mathrm{cp}$, we determine the excluded volume as $V_\mathrm{cp}=NV_0/\phi_\mathrm{max}$.}}
Finally, we {\revisetwo{obtain the compressive strength of dust aggregates}} as
\begin{eqnarray}
P_\mathrm{comp} &=& \frac{E_\mathrm{roll}}{r_0^3}\left(\frac{NV_0}{V-NV_0/\phi_\mathrm{max}}\right)^3 \nonumber\\
&=& \frac{E_\mathrm{roll}}{r_0^3}\left(\frac{1}{\phi}-\frac{1}{\phi_\mathrm{max}}\right)^{-3} \nonumber\\
&\simeq& 4.7\times10^5\mathrm{\ Pa}\left(\frac{\gamma}{100\mathrm{\ mJ\ m^{-2}}}\right)\left(\frac{r_0}{0.1\mathrm{\ \mu m}}\right)^{-2}\nonumber\\
&&\times\left(\frac{\xi_\mathrm{crit}}{8\textrm{\ \AA}}\right)\left(\frac{1}{\phi}-\frac{1}{\phi_\mathrm{max}}\right)^{-3}.
\label{eq:Pcomp_mod}
\end{eqnarray}
{\revisefour{Equation (\ref{eq:Pcomp_mod}) shows that the compressive strength diverges at $\phi_\mathrm{max}$.}}

To compare our simulation results with the {\revisethree{corrected}} {\revisetwo{analytical}} formula, we {\revisetwo{invert}} Equation (\ref{eq:Pcomp_mod}) {\revisetwo{into $\phi$ as a function of}} $P$ {\revisefour{because the input parameter is $P$ and the output parameter is $\phi$ in the applicative situations, such as experiments}}.
The volume filling factor determined by Equation (\ref{eq:Pcomp_mod}) is given as
\begin{equation}
\phi_\mathrm{comp}=\left(\frac{E_\mathrm{roll}^{1/3}}{r_0P^{1/3}}+\frac{1}{\phi_\mathrm{max}}\right)^{-1}.\label{eq:phi_mod}
\end{equation}
{\revisetwo{We assume $\phi_\mathrm{max}=\sqrt{2}\pi/6=0.74$, which is the volume filling factor of the hexagonal close-packed and face-centered cubic structures.}}
{\revisefour{We also invert Equation (\ref{eq:Pcomp_kataoka}) as
\begin{equation}
\phi_\mathrm{K13}=\frac{r_0P^{1/3}}{E_\mathrm{roll}^{1/3}}\label{eq:phi_kataoka}
\end{equation}
to compare the corrected analytical formula with the previous formula.}}

\begin{figure}[ht!]
\plotone{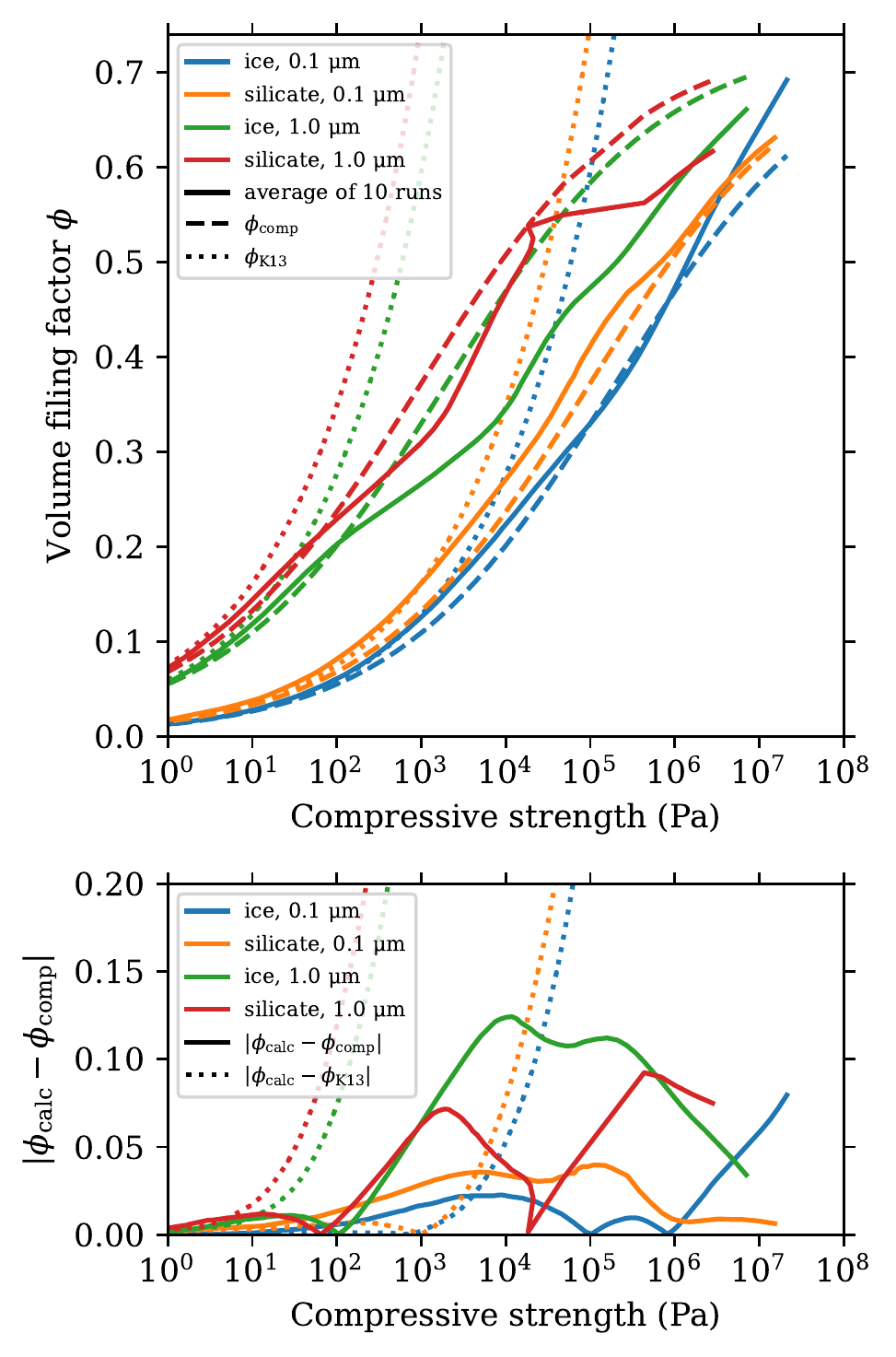}
\caption{{\it Top}: volume filling factor against compressive strength of dust aggregates with different monomer radii and materials.
The colors show 0.1-$\mathrm{\mu m}$-radius ice (blue), 1.0-$\mathrm{\mu m}$-radius ice (green), 0.1-$\mathrm{\mu m}$-radius silicate (orange), and 1.0-$\mathrm{\mu m}$-radius silicate (red).
The parameters are listed in Table \ref{tab:parameters}.
{\revisefour{The solid, dashed, and dotted lines show the calculated volume filling factor $\phi_\mathrm{calc}$, the corrected analytical formula $\phi_\mathrm{comp}$ (Equation (\ref{eq:phi_mod})), and the previous formula $\phi_\mathrm{K13}$ (Equation (\ref{eq:phi_kataoka})), respectively.}}
{\it Bottom}: absolute error between the calculated volume filling factor and that of the corrected analytical formula {\revisefour{(solid) and between the calculated volume filling factor and that of the previous formula (dotted)}}.
\label{fig:comp_modify}}
\end{figure}

{\revisefive{We confirm that the corrected analytical formula is a much better approximation than the previous formula.}}
In the top panel of Figure \ref{fig:comp_modify}, we plot the corrected analytical formula for the volume filling factor {\revisefour{$\phi_\mathrm{comp}$}} (Equation (\ref{eq:phi_mod})){\revisefour{, the previous formula $\phi_\mathrm{K13}$ (Equation (\ref{eq:phi_kataoka})),}} and the calculated volume filling factors of numerical simulations.
{\revisefour{We plot them for $P\geq1\mathrm{\ Pa}$ to focus on when $\phi>0.1$.}}
We note that monomers in our simulations can deform elastically, so that {\revisefour{the volume filling factor}} {\revisefive{could be}} higher than that of the hexagonal close-packed and face-centered cubic structures.
In the bottom panel of Figure \ref{fig:comp_modify}, we plot errors between the calculated {\reviseone{volume filling factor}} and the {\revisethree{corrected}} {\revisetwo{analytical}} formula {\revisefour{and between the calculated volume filling factor and the previous formula}}, which {\revisefour{are}} given as {\reviseone{$|\phi_\mathrm{calc}-\phi_\mathrm{comp}|$}} {\revisefour{and $|\phi_\mathrm{calc}-\phi_\mathrm{K13}|$, respectively}}.
{\revisetwo{The absolute}} errors are smaller than {\reviseone{0.1}} {\revisetwo{for most cases}} {\revisefour{of the corrected analytical formula}}.
{\revisefive{The error would be due to the complexity of numerical simulations, for example, monomers can deform elastically.}}

\subsection{{\revisefive{Monomer Disruption}}}
\label{subsec:discuss:monomerdisruption}

{\revisefive{In this section, we discuss the range of the volume filling factor for which compressive strength can be applied.
If the compressive strength is too high, monomers can be broken, and then it could be different from our results.}}

{\revisefour{The strength that materials can be broken has been investigated in the context of material science.
For example, ice can be broken at 5--25 MPa when the temperature is from $-10^\circ$C to $-20^\circ$C \citep{Haynes1978,Petrovic2003}.}}
{\revisefive{Silica glasses, on the other hand, can be broken at $\sim5$ GPa at room temperature \citep{Proctor1967,Bruckner1970,Kurkjian2003}.}}

{\revisefive{First of all, we note that the stress applied to the contact surface between monomers can be higher than the compressive strength because the stress is concentrated on the area of the contact surface $a^2\ll r_0^2$, where $a$ is the radius of the contact surface.
We assume this stress by assuming the equilibrium radius of the contact surface given by \citet{Wada2007} as
\begin{eqnarray}
a_0 &=& \left(\frac{9\pi\gamma r_0^2}{4E^\ast}\right)^{1/3}\nonumber\\
&\simeq& 0.012\mathrm{\ \mu m}\left(\frac{\gamma}{100\mathrm{\ mJ\ m^{-2}}}\right)^{1/3}\nonumber\\
&&\times\left(\frac{r_0}{0.1\mathrm{\ \mu m}}\right)^{2/3}\left(\frac{E^\ast}{3.7\mathrm{\ GPa}}\right)^{-1/3}.
\end{eqnarray}
The stress applied to the contact surface is $(a_0/r_0)^2$ times as large as the compressive strength.}}

\begin{figure}[ht!]
\plotone{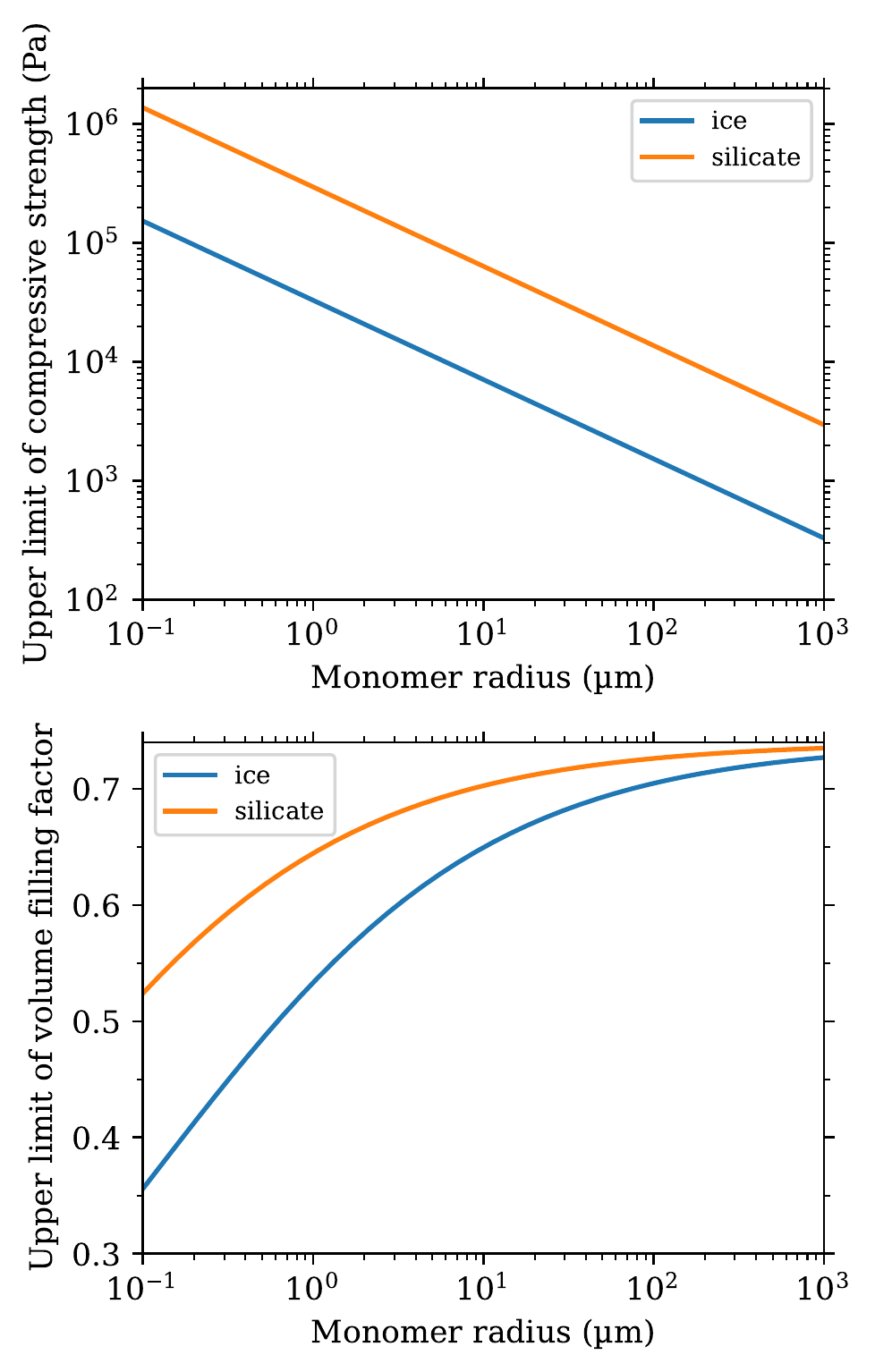}
\caption{{\revisefive{{\it Top}: upper limit of compressive strength for which monomer disruption can be neglected as a function of monomer radius (Equation (\ref{eq:P_upperlimit})).
The blue and orange lines are for ice and silicate, respectively.
The parameters are listed in Table \ref{tab:parameters}.
{\it Bottom}: upper limit of volume filling factor for which monomer disruption can be neglected as a function of monomer radius (Equation (\ref{eq:phi_upperlimit})).}}
\label{fig:upperlimit}}
\end{figure}

{\revisefive{By considering both the strength that materials can be broken and the stress applied to the contact surface, we can estimate the upper limit to which the compressive strength formula (Equation (\ref{eq:Pcomp_mod})) can be applied.
We can estimate the upper limit of the compressive strength as
\begin{eqnarray}
P_\mathrm{ul} &\sim& P_\mathrm{dis}\left(\frac{a_0}{r_0}\right)^2\nonumber\\
&\simeq& 0.014P_\mathrm{dis}\left(\frac{\gamma}{100\mathrm{\ mJ\ m^{-2}}}\right)^{2/3}\nonumber\\
&&\times\left(\frac{r_0}{0.1\mathrm{\ \mu m}}\right)^{-2/3}\left(\frac{E^\ast}{3.7\mathrm{\ GPa}}\right)^{-2/3},
\label{eq:P_upperlimit}
\end{eqnarray}
where $P_\mathrm{dis}$ is the strength that materials can be disrupted.
Here, we assume $P_\mathrm{dis}=10$ MPa and 1 GPa for ice and silicate, respectively.
Then, we can also estimate the upper limit of the volume filling factor as
\begin{equation}
\phi_\mathrm{ul}=\left(\frac{E_\mathrm{roll}^{1/3}}{r_0P_\mathrm{ul}^{1/3}}+\frac{1}{\phi_\mathrm{max}}\right)^{-1}.
\label{eq:phi_upperlimit}
\end{equation}
We plot the upper limit of both the compressive strength (Equation (\ref{eq:P_upperlimit})) and the volume filling factor (Equation (\ref{eq:phi_upperlimit})) in Figure \ref{fig:upperlimit}.
The upper limit of the volume filling factor depends on the monomer radius as well as the material, we find that the upper limits are $\sim0.36$, $\sim0.53$, $\sim0.53$, and $\sim0.64$ in the case of ice $0.1\mathrm{\ \mu m}$, ice $1.0\mathrm{\ \mu m}$, silicate $0.1\mathrm{\ \mu m}$, and silicate $1.0\mathrm{\ \mu m}$, respectively.}}

\subsection{Energy Dissipation Mechanisms} \label{subsec:discuss:energy}

\begin{figure*}[ht!]
\plotone{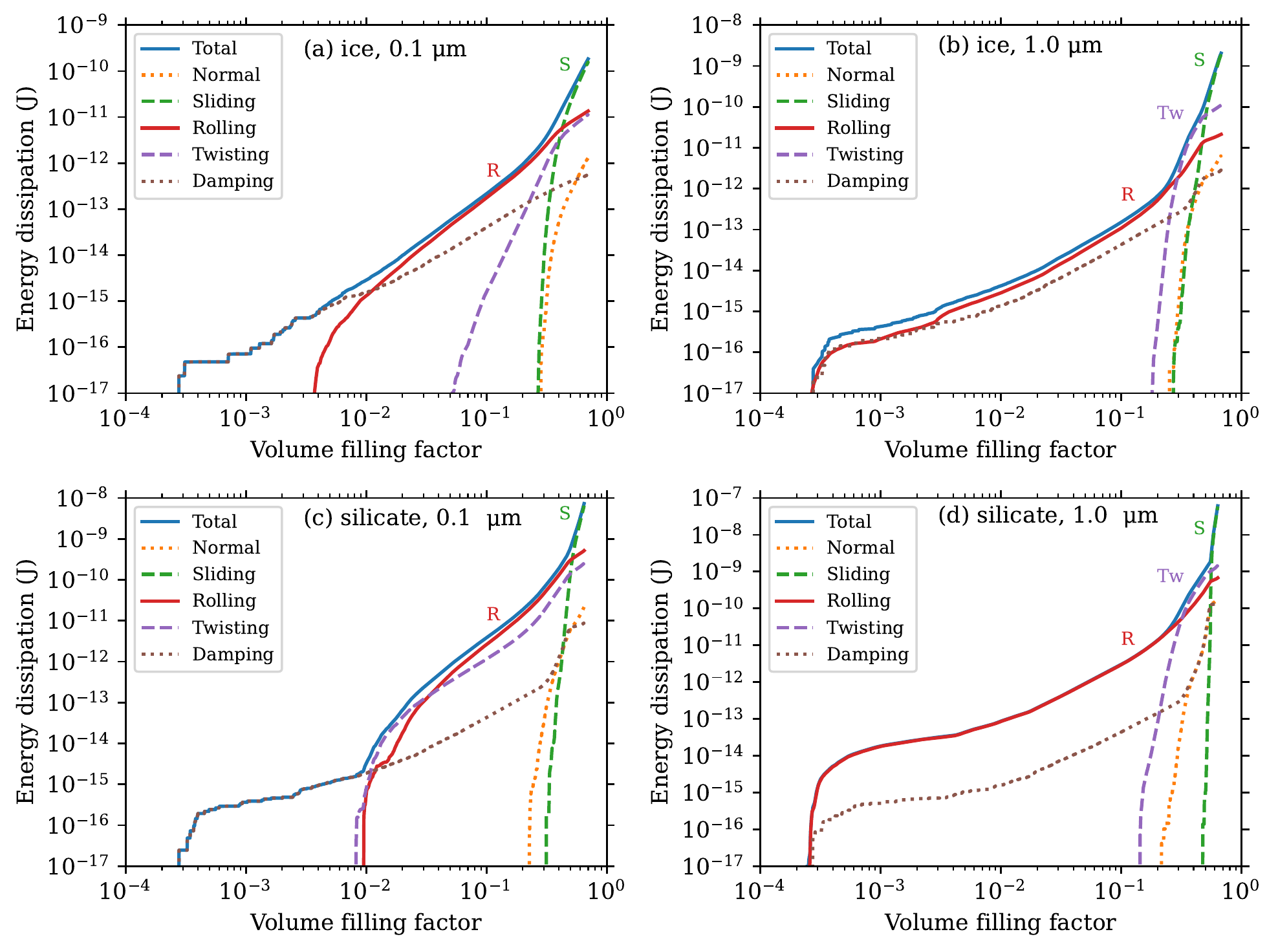}
\caption{Energy dissipations of (a) 0.1-$\mathrm{\mu m}$-radius ice, (b) 1.0-$\mathrm{\mu m}$-radius ice, (c) 0.1-$\mathrm{\mu m}$-radius silicate, and (d) 1.0-$\mathrm{\mu m}$-radius silicate.
The energy dissipation mechanisms are the normal (orange dotted), sliding (green dashed), rolling (red solid), twisting (purple dashed), damping motions (brown dotted), and the total of all energy dissipations (blue solid).
The parameters are listed in Table \ref{tab:parameters}.
\label{fig:energy_material}}
\end{figure*}

% fiducial run: energy dissipation
{\revisefive{We find that the twisting and sliding motions dominate for high volume filling factors ($\phi>0.3$), while the dominant energy dissipation mechanism is the rolling motion for low volume filling factors ($10^{-2}<\phi<0.3$).
This is shown in Figure \ref{fig:energy_material}, where we plot all energy dissipations of the four cases (ice $0.1\mathrm{\ \mu m}$, ice $1.0\mathrm{\ \mu m}$, silicate $0.1\mathrm{\ \mu m}$, and silicate $1.0\mathrm{\ \mu m}$) listed in Table \ref{tab:parameters}.}}
{\revisetwo{The reason why the twisting and sliding motions dominate for high volume filling factors is as follows.
The coordination number defined as the average number of connections per monomer of initial BCCAs is $\simeq2$.
It increases toward $\simeq3$ for high volume filling factors ($\phi>0.3$) \citep[see Figure 3 in][]{Arakawa2019}.
For such a large coordination number, monomers are hard to roll and have to twist and slide to increase the volume filling factor.}}

% damping energy dissipation: wrong compressive strength
{\reviseone{In some cases, the energy dissipation due to damping motion dominates for very low volume filling factors ($\phi<10^{-2}$).
We cannot obtain the correct compressive strength when damping motion dominates.
However, this is not a problem because we focus on the difference between compressive strengths when $\phi\lesssim0.1$ and $\phi>0.1$ in this work.}}

\subsection{Comparison with Previous Studies} \label{subsec:discuss:compare}

% summary of previous studies
There are several experimental and numerical studies about the quasi-static compressive strength of spherical {\revisetwo{silicate and SiO$_2$}} aggregates {\reviseone{with}} volume filling {\reviseone{factors}} higher than 0.1 {\reviseone{\citep[e.g.,][]{Guttler2009,Seizinger2012,Omura2017,Omura2018,Omura2021}}}.
{\reviseone{\citet{Seizinger2012} performed numerical simulations, in which they prepared a {\revisetwo{silicate}} aggregate {\revisetwo{with a monodisperse monomer size distribution}} enclosed in a box with fixed {\revisetwo{boundaries}} on all sides, and moved the top {\revisetwo{boundary}} downwards to mimic the experiments of \citet{Guttler2009}.}}
They used the fitting formula of {\reviseone{volume filling factor $\phi_\mathrm{G09}$}} obtained by \citet{Guttler2009} given as
\begin{equation}
{\reviseone{\phi_\mathrm{G09}}} = \phi_2-\frac{\phi_2-\phi_1}{\exp\left[(\log_{10} P-\log_{10} p_\mathrm{m})/\Delta\right]+1},
\label{eq:Guttler2009_P}
\end{equation}
where $\phi_1=0.15$, $\phi_2=0.58$, $p_\mathrm{m}=16.667$ kPa, and $\Delta=0.562$.
Recently, \citet{Omura2021} fitted the experimental results of \citet{Omura2017,Omura2018}.
{\revisetwo{They used the same experimental setups as \citet{Guttler2009}, but SiO$_2$ aggregates with a polydisperse size distribution.}}
\citet{Omura2021} used the polytropic relationship given as
\begin{equation}
P = K_\mathrm{p}\rho^{(n_\mathrm{p}+1)/n_\mathrm{p}} = K'_\mathrm{p}\phi^{(n_\mathrm{p}+1)/n_\mathrm{p}},
\label{eq:Omura_poly_P}
\end{equation}
where $K_\mathrm{p}$ and $K'_\mathrm{p}$ are constants, $\rho$ is the density, and $n_\mathrm{p}$ is the polytropic index.
{\reviseone{To obtain the volume filling factor as a function of compressive strength, we {\revisetwo{invert}} Equation (\ref{eq:Omura_poly_P}) as
\begin{equation}
\phi_\mathrm{O21} = \left(\frac{P}{K_\mathrm{p}'}\right)^{n_\mathrm{p}/(n_\mathrm{p}+1)}.
\label{eq:Omura_poly}
\end{equation}}}
Their fitting results are listed in Table \ref{tab:Omura2021}.

\begin{figure*}[ht!]
\plotone{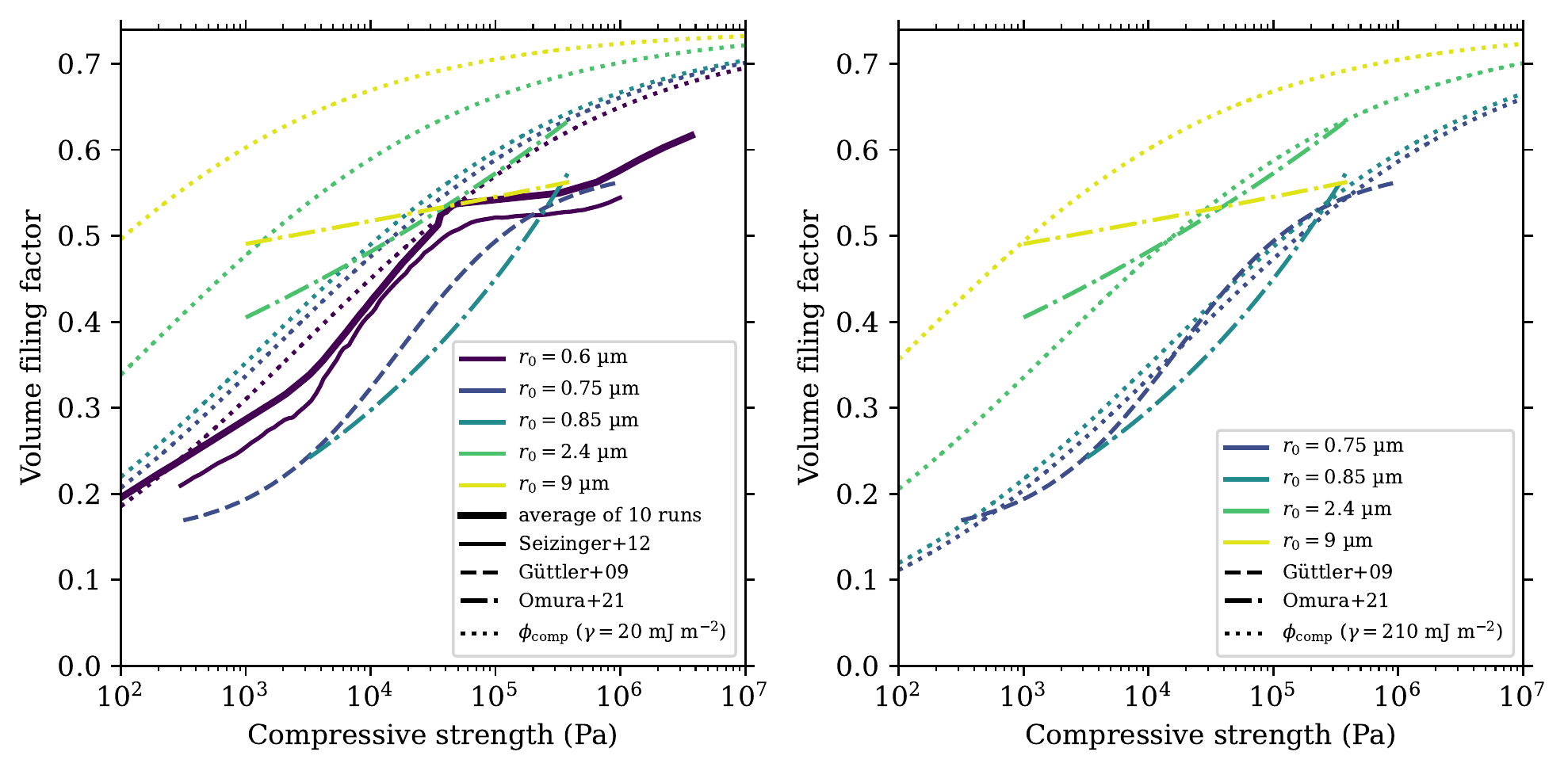}
\caption{{\revisetwo{{\it Left}: volume filling factor against compressive strength of this work and previous studies.
The monomer radii are $r_0=0.6$, 0.75, 0.85, 2.4, and 9 $\mathrm{\mu m}$ from dark to light colors {\revisefour{in all line types}}.
The thick and thin solid lines show the numerical results of silicate aggregates of this work and \citet{Seizinger2012}, respectively.
The other parameters of silicate are listed in Table \ref{tab:parameters}, but we adopt that $C_\mathrm{v}=3\times10^{-7}$.
The dashed and dash-dotted lines show the experimental results of SiO$_2$ aggregates with a monodisperse \citep{Guttler2009} and polydisperse monomer size distribution \citep{Omura2021}, respectively.
We use Equation (\ref{eq:Guttler2009_P}) and Equation (\ref{eq:Omura_poly}) as the fitting results of \citet{Guttler2009} and \citet{Omura2021}, respectively.
We also use the values listed in Table \ref{tab:Omura2021}.
The dotted lines show the analytical formula $\phi_\mathrm{comp}$ (Equation (\ref{eq:phi_mod})) of silicate aggregates.
{\it Right}: the dotted lines are the same as those of the left panel, but we change the surface energy to be $\gamma=210\mathrm{\ mJ\ m^{-2}}$.}}
\label{fig:comp_high_compare}}
\end{figure*}

{\revisetwo{To compare our results with previous ones, we plot them in Figure \ref{fig:comp_high_compare}.}}
We use the fitting formulas of experiments: Equation (\ref{eq:Guttler2009_P}) and Equation (\ref{eq:Omura_poly}).
{\reviseone{We compare our results with the previous numerical study \citep{Seizinger2012}, the previous experimental study of {\revisetwo{SiO$_2$}} aggregates with a monodisperse monomer size distribution \citep{Guttler2009}, and then previous experimental studies with a polydisperse size distribution \citep{Omura2021}.}}

% compare: Seizinger
{\revisetwo{In the left panel of Figure \ref{fig:comp_high_compare},}} our results are in good agreement with previous numerical results.
However, there is a little difference between them because of the different compression setups: to move only the top {\revisetwo{boundary}} \citep{Seizinger2012} or all {\revisetwo{boundaries}} (this work).
We find that this difference in compression setups is negligible.

% compare: Guttler
In the case of the previous experimental study of {\revisetwo{SiO$_2$}} aggregates with a monodisperse monomer size distribution \citep{Guttler2009}, {\revisetwo{the left panel of Figure \ref{fig:comp_high_compare} shows that}} there is a discrepancy between their results and our analytical {\revisetwo{formula}} $\phi_\mathrm{comp}$.
{\revisetwo{This discrepancy may arise from the difference in surface energy.
In our simulations, we assume that the surface energy of silicate (SiO$_2$) is $\gamma=20\mathrm{\ mJ\ m^{-2}}$, but}} some studies suggest that {\revisetwo{it}} can be higher \citep[e.g.,][]{Yamamoto2014,Kimura2015}.
Therefore, we {\revisetwo{search for the best-fitted surface energy and find that $\phi_\mathrm{comp}$ with $\gamma\simeq210\pm90\mathrm{\ mJ\ m^{-2}}$ is in good agreement with the experimental results of \citet{Guttler2009} as shown in the right panel of Figure \ref{fig:comp_high_compare}.}}

% compare: Omura
{\revisetwo{We also explain the previous experimental results of SiO$_2$ aggregates with a polydisperse monomer size distribution \citep{Omura2021} by assuming a higher surface energy than $\gamma=20\mathrm{\ mJ\ m^{-2}}$.
The left panel of Figure \ref{fig:comp_high_compare} shows that there is a discrepancy between their results and our analytical formula $\phi_\mathrm{comp}$ when $\gamma=20\mathrm{\ mJ\ m^{-2}}$, while the right panel shows that their results are in good agreement with $\phi_\mathrm{comp}$ when $\gamma=210\mathrm{\ mJ\ m^{-2}}$.
However, there remains a discrepancy, especially for a larger monomer radius.}}
When the monomer size distribution is polydisperse, larger monomers get stuck at first {\revisetwo{during aggregate compression}}, and then smaller monomers get stuck.
{\revisetwo{We interpret this discrepancy as the uncertainty of the volume filling factor of realistic dust aggregates which have a monomer size distribution.}}

\begin{deluxetable*}{lccccc}
\tablecaption{Fitting results of compression experiments of \citet{Omura2021}
\label{tab:Omura2021}}
\tablewidth{0pt}
\tablehead{
\colhead{Name} & \colhead{Median radius ($\mathrm{\mu m}$)} & \colhead{$K'_\mathrm{p}$ (Pa)} & \colhead{$n_\mathrm{p}$} &
\multicolumn{2}{c}{Fitting range} \\
\cline{5-6}
\colhead{} & \colhead{} & \colhead{} &
\colhead{} & \colhead{Min (Pa)} & \colhead{Max (Pa)}
}
\startdata
Silica beads & 0.85 & $(8.35\pm0.23)\times10^6$ & $(2.201\pm0.013)\times10^{-1}$ & $3.2\times10^3$ & $4.0\times10^5$ \\
Fly ash & 2.4 & $(1.649\pm0.025)\times10^8$ & $(8.134\pm0.014)\times10^{-2}$ & $1.0\times10^3$ & $3.9\times10^5$ \\
Glass beads & 9 & $(2.951\pm0.093)\times10^{16}$ & $(2.3496\pm0.0027)\times10^{-2}$ & $1.0\times10^3$ & $3.9\times10^5$ \\
\enddata
% \tablecomments{}
\end{deluxetable*}

\section{Conclusions} \label{sec:conclusion}

We {\revisetwo{performed numerical simulations of the compression of dust aggregates}} and formulated the compressive strength {\revisetwo{that can treat a full range of}} volume filling factors.
{\reviseone{We used a}} monomer interaction model based on \citet{Dominik1997} and \citet{Wada2007}.
In our simulations, we created a BCCA at first and compressed it {\reviseone{sufficiently slowly}} and three-dimensionally by moving periodic boundaries.
{\revisetwo{We calculated the compressive strength by using the translational kinetic energy and the sum of the forces acting at all connections per unit volume.}}
For every parameter set, we performed 10 simulations with different initial BCCAs {\reviseone{and took an average of them}}.

Our main findings of the compressive strength of dust aggregates are as follows.

\begin{enumerate}
\item As a result of numerical simulations, we found that the compressive strength {\revisetwo{becomes sharply harder when the volume filling factor exceeds 0.1.}}
We also found that the compressive strength for high volume filling factors ($\phi\gtrsim0.3$) does not depend on the critical rolling displacement.

\item We {\revisethree{corrected}} the analytical formula of compressive strength {\revisetwo{by taking into account the closest packing of aggregates for high volume filling factors.}}
{\reviseone{The {\revisethree{corrected}} formula is {\revisethree{given by Equation (\ref{eq:Pcomp_mod}) in Section \ref{subsec:discuss:formulate}}} as follows:}}
\begin{eqnarray}
P_\mathrm{comp} &=& \frac{E_\mathrm{roll}}{r_0^3}\left(\frac{1}{\phi}-\frac{1}{\phi_\mathrm{max}}\right)^{-3}\nonumber\\
&\simeq& 4.7\times10^5\mathrm{\ Pa}\left(\frac{\gamma}{100\mathrm{\ mJ\ m^{-2}}}\right)\left(\frac{r_0}{0.1\mathrm{\ \mu m}}\right)^{-2}\nonumber\\
&&\times\left(\frac{\xi_\mathrm{crit}}{8\textrm{\ \AA}}\right)\left(\frac{1}{\phi}-\frac{1}{\phi_\mathrm{max}}\right)^{-3},
\end{eqnarray}
where $E_\mathrm{roll}$ is the energy needed {\reviseone{for a monomer}} to roll a distance of $(\pi/2)r_0$ (Equation (\ref{eq:E_roll})), $r_0$ is the monomer radius, $\phi_\mathrm{max}=\sqrt{2}\pi/6=0.74$ is the volume filling factor of the closest packing{\reviseone{, $\gamma$ is the surface energy, and $\xi_\mathrm{crit}$ is the critical rolling displacement.
We confirmed that the {\revisethree{corrected}} {\revisetwo{analytical}} formula reproduces our simulation results including parameter dependences.}}
{\revisefive{In terms of the monomer disruption, the corrected formula is valid for $\phi\lesssim0.36$, 0.53, 0.53, and 0.64 in the case of 0.1-$\mathrm{\mu m}$-radius ice, 1.0-$\mathrm{\mu m}$-radius ice, 0.1-$\mathrm{\mu m}$-radius silicate, and 0.1-$\mathrm{\mu m}$-radius silicate monomers, respectively.}}

\item We found that {\reviseone{twisting and sliding motions dominate for high volume filling factors {\revisetwo{($\phi>0.3$)}}, while rolling motion dominates for low volume filling factors {\revisetwo{($\phi<0.3$)}}.
We explained the reason {\revisetwo{why}} the twisting and sliding motions dominate by the increase of coordination number.}}

\item Our numerical results are consistent with the previous numerical results \citep{Seizinger2012}.
However, there is a discrepancy between the previous experimental results \citep{Guttler2009} and our analytical {\revisetwo{formula}}.
We found that our analytical {\revisetwo{formula is}} consistent with the experimental results if we assume the surface energy of silicate {\revisetwo{is $\simeq210\pm90\mathrm{\ mJ\ m^{-2}}$}}.

\end{enumerate}

\begin{acknowledgments}

This work was supported by JSPS KAKENHI Grant Numbers JP19J20351 and JP22J00260.
This work has made use of NASA's Astrophysics Data System.
This work has made use of adstex (\url{https://github.com/yymao/adstex}).

\end{acknowledgments}


\software{VisIt \citep{VisIt}}

\bibliography{paper}{}
\bibliographystyle{aasjournal}
% https://github.com/yymao/adstex

\appendix

\section{Derivation of Compressive Strength} \label{apsec:compstrength}

{\revisetwo{In this appendix, we explain the detailed derivation of compressive strength related to Section \ref{subsec:setting:measure}.}}

{\revisetwo{The equation of motion of monomer $i$ is given as
\begin{equation}
m_0\frac{\mathrm{d}^2\bm{x}_i}{\mathrm{d}t^2}=\bm{W}_i+\bm{F}_i,
\label{eq:EOM_comp}
\end{equation}
where $\bm{W}_i$ is the force exerted from the computational boundaries on monomer $i$ and $\bm{F}_i$ is the total force from other monomers on monomer $i$.
The compressive strength relates to $\bm{W}_i$.}}

{\revisetwo{To describe the first term on the right-hand side of Equation (\ref{eq:EOM_comp}) by using the compressive strength, we take an inner product of $\bm{x}_i$ and Equation (\ref{eq:EOM_comp}), and take a long-time average with time interval $\tau$.
The left-hand side of Equation (\ref{eq:EOM_comp}) becomes
\begin{equation}
\frac{m_0}{\tau}\int_0^\tau \bm{x}_i\cdot\frac{\mathrm{d}^2\bm{x}_i}{\mathrm{d}t^2}\mathrm{d}t=\frac{m_0}{\tau}\left[\bm{x}_i\cdot\frac{\mathrm{d}\bm{x}_i}{\mathrm{d}t}\right]_0^\tau-\frac{m_0}{\tau}\int_0^\tau\frac{\mathrm{d}\bm{x}_i}{\mathrm{d}t}\cdot\frac{\mathrm{d}\bm{x}_i}{\mathrm{d}t}\mathrm{d}t.
\label{eq:EOMleft_comp}
\end{equation}
The first term on the right-hand side of Equation (\ref{eq:EOMleft_comp}) vanishes when $\tau\to\infty$.
Writing a long-time average as $\langle\rangle_t$ and summing Equation (\ref{eq:EOM_comp}) over all $i$, we have
\begin{equation}
\left\langle\sum_{i=1}^N\frac{m_0}{2}\left(\frac{\mathrm{d}\bm{x}_i}{\mathrm{d}t}\right)^2\right\rangle_t = -\frac{1}{2}\left\langle\sum_{i=1}^N\bm{x}_i\cdot\bm{W}_i\right\rangle_t -\frac{1}{2}\left\langle\sum_{i=1}^N\bm{x}_i\cdot\bm{F}_i\right\rangle_t.
\label{eq:EOM2_comp}
\end{equation}
The left-hand side of Equation (\ref{eq:EOM2_comp}) is the time-averaged kinematic energy of all monomers defined as
\begin{equation}
K = \left\langle\sum_{i=1}^N\frac{m_0}{2}\left(\frac{\mathrm{d}\bm{x}_i}{\mathrm{d}t}\right)^2\right\rangle_t.
\end{equation}
The first term on the right-hand side of Equation (\ref{eq:EOM2_comp}) relates to the compressive strength $P_\mathrm{calc}$.
The force on the {\revisefive{computational}} boundary of the area $\mathrm{d}S_\mathrm{b}$ is $P_\mathrm{calc}\bm{n}_\mathrm{b}\mathrm{d}S_\mathrm{b}$, where $\bm{n}_\mathrm{b}$ is the normal vector of the boundary directed outward.
Then,
\begin{eqnarray}
\left\langle\sum_{i=1}^N\bm{x}_i\cdot\bm{W}_i\right\rangle_t &=& -\int_{S_\mathrm{b}}P_\mathrm{calc}\bm{n}_\mathrm{b}\cdot\bm{x}\mathrm{d}S_\mathrm{b} \nonumber \\
&=& -P_\mathrm{calc}\int_V\mathrm{div}\bm{x}\mathrm{d}V \nonumber \\
&=& -P_\mathrm{calc}\int_V\left(\frac{\partial x}{\partial x}+\frac{\partial y}{\partial y}+\frac{\partial z}{\partial z}\right)\mathrm{d}V \nonumber \\
&=& -3P_\mathrm{calc}V.
\end{eqnarray}
The total force from other monomers to monomer $i$ can be described as
\begin{equation}
\bm{F}_i = \sum_{j\neq i}\bm{f}_{i,j}.
\label{eq:totalforce}
\end{equation}
Equations (\ref{eq:EOM2_comp})--(\ref{eq:totalforce}) yield
\begin{equation}
P_\mathrm{calc} = \frac{2K}{3V} + \frac{1}{3V}\left\langle\sum_{i<j}(\bm{x}_i-\bm{x}_j)\cdot\bm{f}_{i,j}\right\rangle_t
\label{eq:P_compfinal}
\end{equation}
because of the relation that $\bm{f}_{i,j}=-\bm{f}_{j,i}$.}}

\section{{\revisefour{Other Parameter Dependences}}}
\label{apsec:parameterdepend}

{\revisefour{In this appendix, we show dependences on the number of monomers $N$, the strain rate parameter $C_\mathrm{v}$, the damping coefficient $k_\mathrm{n}$, and the time-step.}}

\begin{figure}[ht!]
\plotone{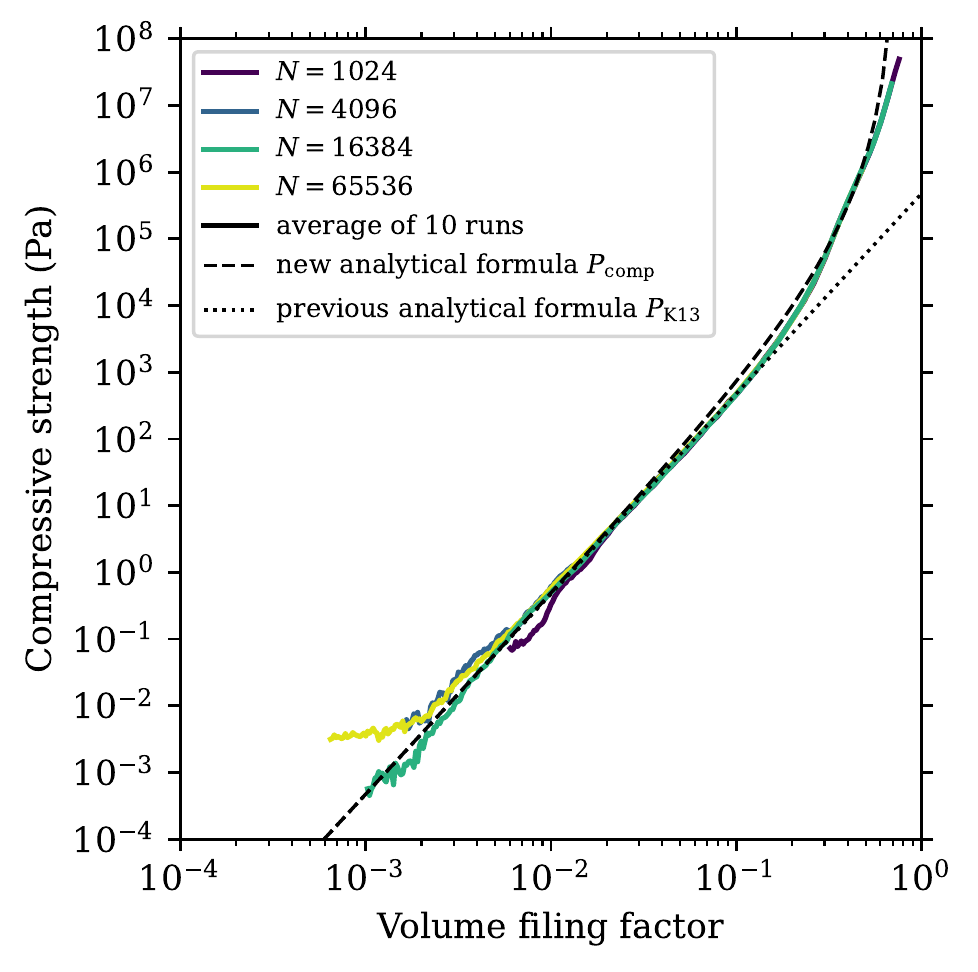}
\caption{{\revisefour{Compressive strength against volume filling factor of dust aggregates that contain ice monomers of 0.1-$\mathrm{\mu m}$ radius with different numbers of monomers $N$.
The numbers of monomers are $N=1024$, 4096, 16384, and 65536 from dark to light colors.
We adopt $C_\mathrm{v}=3\times10^{-7}$ for $N=1024$, 4096, and 65536 because of the calculation cost.
The other parameters are fiducial values in Table \ref{tab:parameters}.
The solid, dashed, and dotted lines show the averages of the 10 runs, our corrected analytical formula (Equation (\ref{eq:Pcomp_mod})), and the analytical formula (Equation (\ref{eq:Pcomp_kataoka})) of \citet{Kataoka2013}, respectively.}}
\label{fig:comp_num}}
\end{figure}

{\revisefive{First, we confirm that there is no dependence on the number of monomers, i.e., the size of the calculation box.
This is shown in Figure \ref{fig:comp_num}, where we plot compressive strength when $N=1024$, 4096, 16384, and 65536.}}

\begin{figure}[ht!]
\plotone{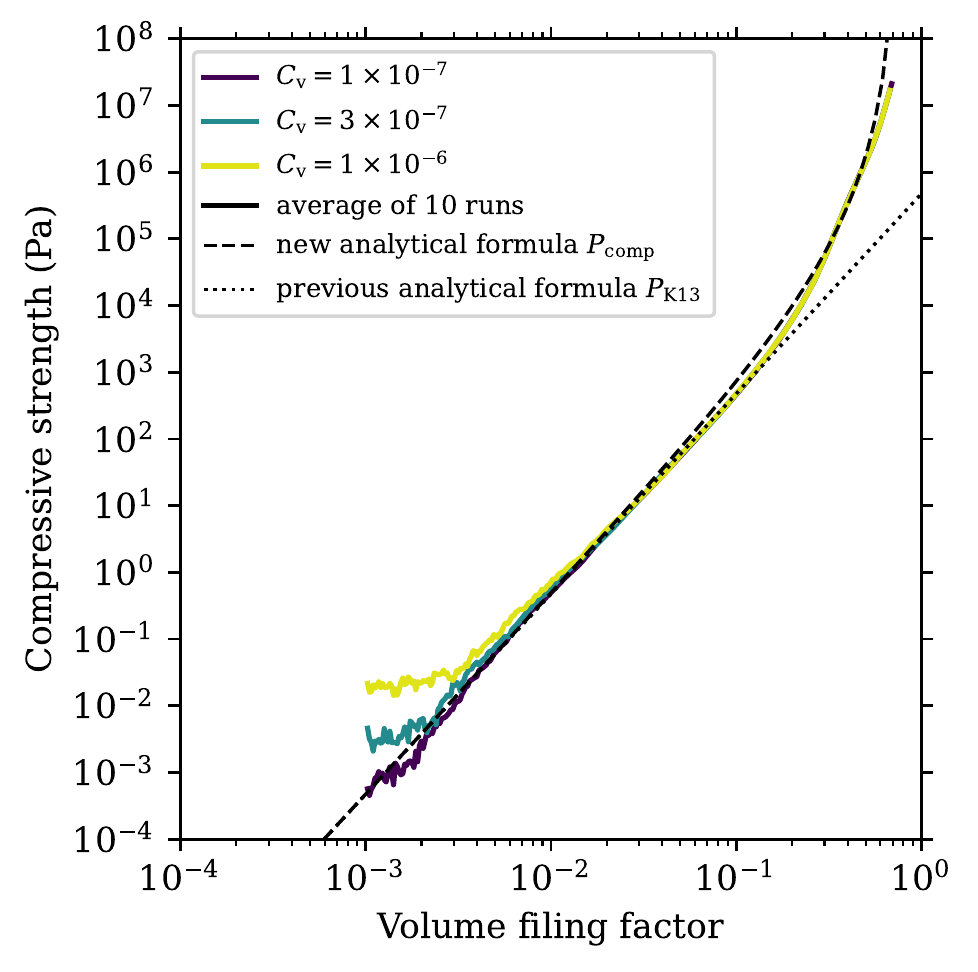}
\caption{{\revisefour{Compressive strength against volume filling factor of dust aggregates that contain ice monomers of 0.1-$\mathrm{\mu m}$ radius with different strain rate parameters $C_\mathrm{v}$.
The strain rate parameters are $C_\mathrm{v}=1\times10^{-7}$, $3\times10^{-7}$, and $1\times10^{-6}$ from dark to light colors.
The other parameters are fiducial values in Table \ref{tab:parameters}.
The solid, dashed, and dotted lines show the averages of the 10 runs, our corrected analytical formula (Equation (\ref{eq:Pcomp_mod})), and the analytical formula (Equation (\ref{eq:Pcomp_kataoka})) of \citet{Kataoka2013}, respectively.}}
\label{fig:comp_vel}}
\end{figure}

{\revisefive{Second, we verify that the strain rate parameter, which refers to the velocity at the computational boundaries, does not exhibit any dependence.
This is shown in Figure \ref{fig:comp_vel}, where we plot compressive strength when $C_\mathrm{v}=1\times10^{-7}$, $3\times10^{-7}$, and $1\times10^{-6}$.}}
{\revisefour{There are fluctuations of compressive strength when $\phi\lesssim3\times10^{-3}$ because dust aggregates are not attached to all {\revisefive{computational}} boundaries.
We note that compressive strength in this work is quasi-static, so it does not depend on the velocity at the {\revisefive{computational}} boundaries}} {\revisefive{if it is small enough.}}

\begin{figure}[ht!]
\plotone{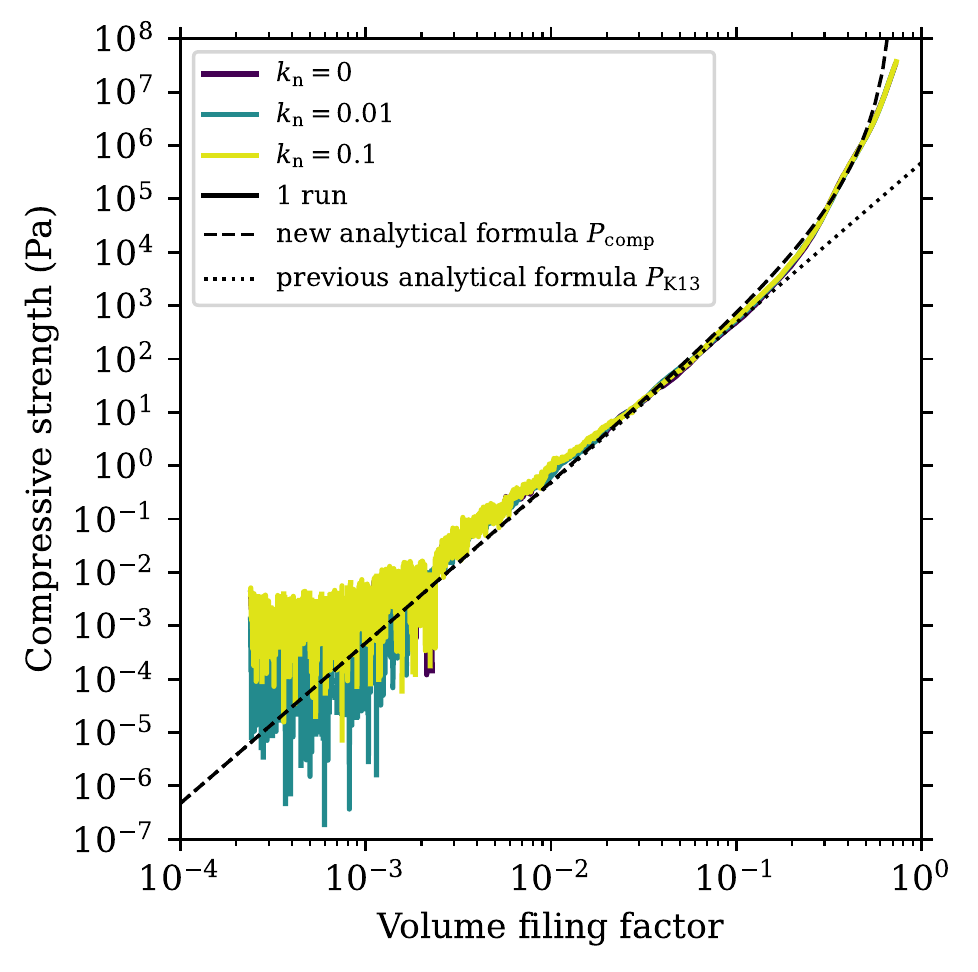}
\caption{{\revisefour{Compressive strength against volume filling factor of dust aggregates that contain ice monomers of 0.1-$\mathrm{\mu m}$ radius with different damping coefficients $k_\mathrm{n}$.
The damping coefficients are $k_\mathrm{n}=0$, 0.01, and 0.1 from dark to light colors.
We adopt $C_\mathrm{v}=3\times10^{-7}$ for $k_\mathrm{n}=0$ and 0.1 because of the calculation cost.
The other parameters are fiducial values in Table \ref{tab:parameters}.
The solid, dashed, and dotted lines show the result of a run, our corrected analytical formula (Equation (\ref{eq:Pcomp_mod})), and the analytical formula (Equation (\ref{eq:Pcomp_kataoka})) of \citet{Kataoka2013}, respectively.}}
\label{fig:comp_damp}}
\end{figure}

{\revisefive{Third, we confirm that the damping coefficient does not exhibit any dependence.
This is shown in Figure \ref{fig:comp_damp}, where we plot compressive strength when $k_\mathrm{n}=0$, 0.01, and 0.1.}}

\begin{figure}[ht!]
\plotone{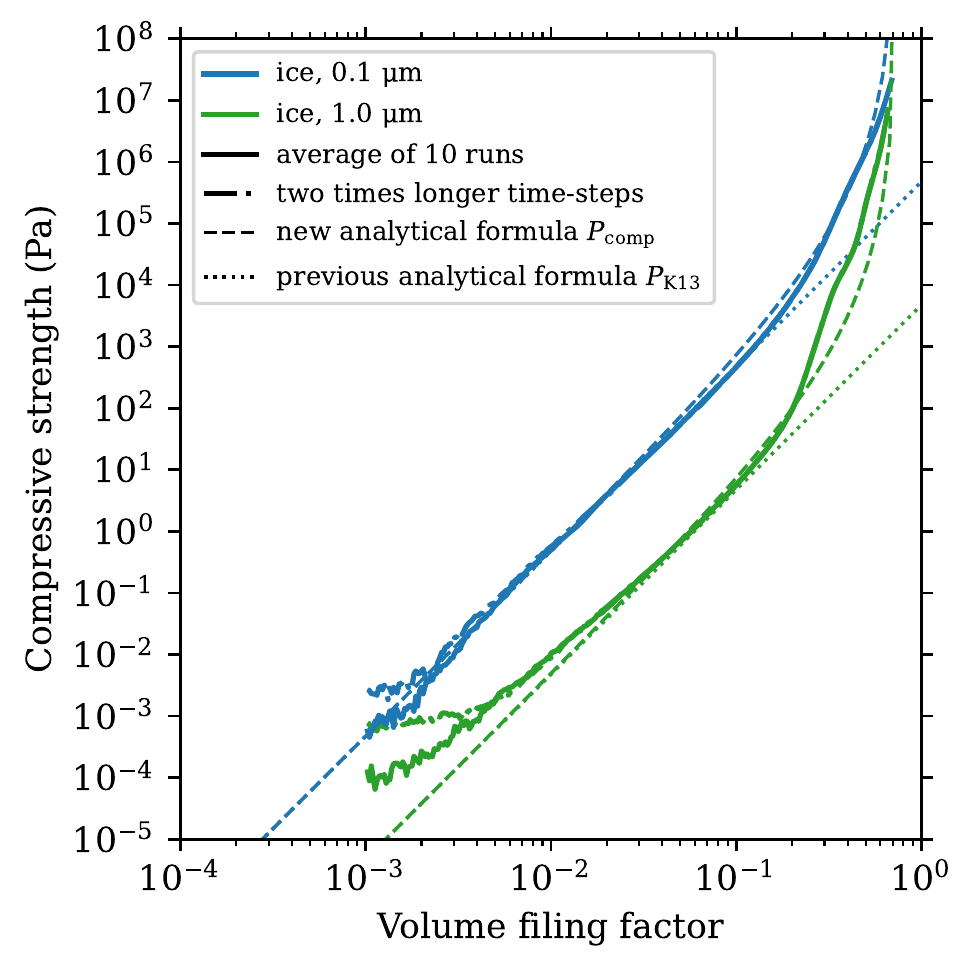}
\caption{{\revisefour{Compressive strength against volume filling factor of dust aggregates that contain ice monomers of 0.1-$\mathrm{\mu m}$ (blue) and 1.0-$\mathrm{\mu m}$ (green) radius with different time-steps.
The other parameters are fiducial values in Table \ref{tab:parameters}.
The solid, dash-dotted, dashed, and dotted lines show the averages of the 10 runs, the averages of the 10 runs with two times longer time-steps, our corrected analytical formula (Equation (\ref{eq:Pcomp_mod})), and the analytical formula (Equation (\ref{eq:Pcomp_kataoka})) of \citet{Kataoka2013}, respectively.
We adopt $C_\mathrm{v}=3\times10^{-7}$ for two times longer time-steps because of the calculation cost.}}
\label{fig:comp_timestep}}
\end{figure}

{\revisefive{Finally, we verify that the results are not affected by the length of the time-step because the compressive strength in this work is quasi-static.
This is shown in Figure \ref{fig:comp_timestep}, where we plot compressive strength in the two cases (ice $0.1\mathrm{\ \mu m}$ and ice $1.0\mathrm{\ \mu m}$) listed in Table \ref{tab:parameters} and when the time-step is two times longer.}}

\end{document}